\documentclass[prl,superscriptaddress,twocolumn,amsfonts,amsmath,a4paper]{revtex4}
\usepackage{graphicx}

\begin{document}

\title{Dynamics of a rod in a homogeneous/inhomogeneous frozen disordered medium:
Correlation functions and non-Gaussian effects}

\author{Angel J. Moreno}
\affiliation{Laboratoire des Verres. Universit\'{e} Montpellier 2. Place
  E. Bataillon.  CC 069. F-34095 Montpellier, France.}
\affiliation{Present address: Donostia International Physics Center, Paseo Manuel de Lardizabal 4,
E-20018 San Sebasti\'{a}n, Spain.} 
\author{Walter Kob}
\affiliation{Laboratoire des Collo\"{i}des, Verres et Nanomat\'{e}riaux, Universit\'{e} Montpellier 2, 
F-34095 Montpellier, France}

\begin{abstract}
\vspace{0.3 cm}
We present molecular dynamics simulations of the motion of a single rigid
rod in a disordered static 2d-array of disk-like obstacles.  Two different
configurations have been used for the latter: A completely random one,
and which thus has an inhomogeneous structure, and an homogeneous ``glassy'' one,
obtained from freezing a liquid of soft disks in equilibrium.  Small
differences are observed between both structures for the translational
dynamics of the rod center-of-mass. In contrast to this, the rotational
dynamics in the glassy host medium is strongly slowed down in comparison
with the random one. We calculate angular correlation functions 
for a wide range of rod length $L$ and density of obstacles $\rho$ as
control parameters. A two-step decay is observed for large values of $L$
and $\rho$, in analogy with supercooled liquids at temperature close to
the glass transition. In agreement with the prediction of the
Mode Coupling Theory, a time-length
and time-density scaling is obtained. In order to get insight on the
relation between the heterogeneity of the dynamics and the structure
of the host medium, we determine the deviations from Gaussianity at different length
scales.
Strong deviations are obtained even at spatial scales much larger than the
rod length.  The magnitude of these deviations is independent of the nature of the host
medium. This result suggests that the large
scale translational dynamics of the rod is affected only weakly by the
presence of inhomogeneities in the host medium.
\\\\
{\bf Ref.:  AIP Conference Proceedings 708 (2004) 576-582}
\end{abstract}

\maketitle


\section{I. Introduction}

Since it was initially introduced by Lorentz as a model for the electrical
conductivity in metals \cite{1}, the problem of the Lorentz gas has
given rise to a substantial theoretical effort aimed to understand its 
properties~\cite{2,3,4,5,6,7,8}.  In this model,
a {\it single} classical particle moves through a disordered array
of {\it static} obstacles.  It can thus be used as a simplified picture of the
motion of a light atom in a disordered environment of heavy particles
having a much slower dynamics.  In the simplest case, where the
diffusing particle and the obstacles are modeled as hard spheres, an
exact solution for the diffusion constant exists in the limit of low
densities of obstacles \cite{9}.  However, the problem becomes highly
non-trivial with increasing density, where dynamic correlations and
memory effects start to become important for the motion of the diffusing
particle, and the system shows the typical features of the dynamics of
supercooled liquids or dense colloidal systems, such as a transition to
a non-ergodic phase of zero diffusivity \cite{2,3,4,6}.  In particular,
diffusion constants and correlation functions are non-analytical functions
of the density \cite{2,4,7,8}. Moreover, correlation functions show
non-exponential long-time decays.

Diffusing particles and obstacles are generally modeled as disks or
spheres in two and three dimensions, respectively. Much less attention has been
paid to systems that have orientational degrees of freedom.  Motivated
by this latter question, we have recently started an investigation,
at low and moderate densities of obstacles, on a generalization of the
Lorentz gas, namely a model in which the diffusing particle is a rigid rod
\cite{10}.  An array of randomly distributed disks has been used for the
host medium.  For simplicity simulations have been done in two dimensions,
reducing the degrees of freedom to the center-of-mass position and the
orientation of the rod axis.  As in the case of supercooled liquids
\cite{11,12,13,14,15} or dense colloidal systems \cite{16,17,18}, one observes at
intermediate times a caging regime for the rod center-of-mass motion, which is due to the
steric hindrance produced by the presence of the neighboring obstacles.

More interestingly, strong deviations from Gaussianity have been obtained
for the incoherent intermediate scattering function at wavelengths much
longer than the rod length, giving evidence for a strongly heterogeneous character
of the long-time dynamics at such length scales.  The inhomogeneous
structure of the model used for the host medium has been pointed out
as a possible origin of such non-Gaussian effects since a random
configuration features large holes on one side but on the other side also 
dense clusters of 
obstacles. The presence of holes might lead to a finite probability of
jumps that are much longer than the average ``jump length'' and thus to a heterogeneous
dynamics.

In order to shed new light on this question, we have repeated the
simulations by taking a disordered but {\it homogeneous} configuration
of the obstacles instead of a random one.  We will see, however, that
large scale non-Gaussian effects are not significantly affected
by the particular choice of the configuration of the obstacles.
Some information on angular correlation functions is also presented.
The article is organized as follows: In Section II we present the
model and give details of the simulation. Translational and angular
mean-squared displacements are presented in Section III.  The behavior
of angular correlation functions is shown and discussed in Section IV,
in terms of the Mode Coupling Theory.  Non-Gaussian effects in the
random and glassy models of the host medium are investigated in Section
V. Conclusions are given in Section VI.

\section{II. Model and details of the simulation}

The rigid rod, of mass $M$, was modeled as $N$ aligned point particles
of equal mass $m=M/N$, with a bond length $2\sigma$. The rod length is
therefore given by $L=(2N-1)\sigma$. The positions of the obstacles in
the random configuration were generated by a standard Poisson process. In
order to obtain the glassy host medium, we equilibrated at a reduced particle
density $\rho = 0.77$ and at temperature $T=\epsilon/k_{\rm B}$ a
two-dimensional array of point particles interacting via a soft-disk
potential $V(r)=\epsilon(\sigma/r)^{12}$. This procedure produced an
homogeneous liquid-like configuration. The latter was then permanently
frozen and was expanded or shrunk to obtain the desired density of
obstacles, defined as $\rho = n_{\rm obs}/l_{\rm box}^{2}$, with $n_{\rm
obs}$ the number of obstacles and $l_{\rm box}$ the length of the square
simulation box used for periodic boundary conditions.

The same soft-disk potential $V(r)$ was used for the interaction between
the particles forming the rod and the obstacles.  For computational
efficiency, $V(r)$ was truncated and shifted at a cutoff distance of
2.5$\sigma$. In the following, space and time will be measured in the
reduced units $\sigma$ and $(\sigma^{2}m/\epsilon)^{1/2}$, respectively.
Typically 600-1000 realizations of the ensemble rod-obstacles were
considered.  The set of rods was equilibrated at $T = \epsilon/k_{\rm
B}$. After the equilibration, a production run was done at constant
energy and results were averaged over the different realizations. These
runs covered $10^{6}$ time units, corresponding to $(1-5)\cdot 10^{8}$
time steps, depending on the step size used for the different rod lengths
and densities.	Run times were significantly longer than the relaxation
times of the system.

\section{III. Mean-squared displacements}

Figs. 1a and 1b show respectively for $\rho =6\cdot 10^{-3}$ a comparison
between the random and the glassy configuration for the mean-squared
displacement of the rod center-of-mass $\langle(\Delta r(t))^{2}\rangle$
and the mean-squared angular displacement $\langle(\Delta
\Phi(t))^{2}\rangle$. Brackets denote ensemble average.  In order to
facilitate the observation of the different dynamic regimes, data have been divided by
the time $t$.  The comparison covers a wide range of rod lengths from
$L \sim 0.1d_{\rm nn}$ to $L \sim 10d_{\rm nn}$, where $d_{\rm nn}
=\rho^{-1/2} \approx 13$, is the average distance between obstacles for
the mentioned density.  At short times the rod does not feel the presence
of the neighboring obstacles and $\langle(\Delta r(t))^{2}\rangle$ and
$\langle(\Delta \Phi(t))^{2}\rangle$ show the quadratic time-dependence
characteristic of a ballistic motion. For the shortest rods a sharp
transition to the long-time linear regime is observed. In contrast to this,
for long rods a crossover regime between both limits, showing a weaker
time dependence that the ballistic motion, is present over 1-2 decades of
intermediate times. Such a crossover corresponds to the well-known caging
regime \cite{11,12,13,14,15,16,17,18} observed in supercooled liquids
or dense colloidal systems.  Due to the presence of the neighboring
obstacles, the particle is trapped within an ``effective cage'' for some
time until it escapes from it and begins to show a diffusive behavior.

\begin{figure}
\includegraphics[width=.87\linewidth]{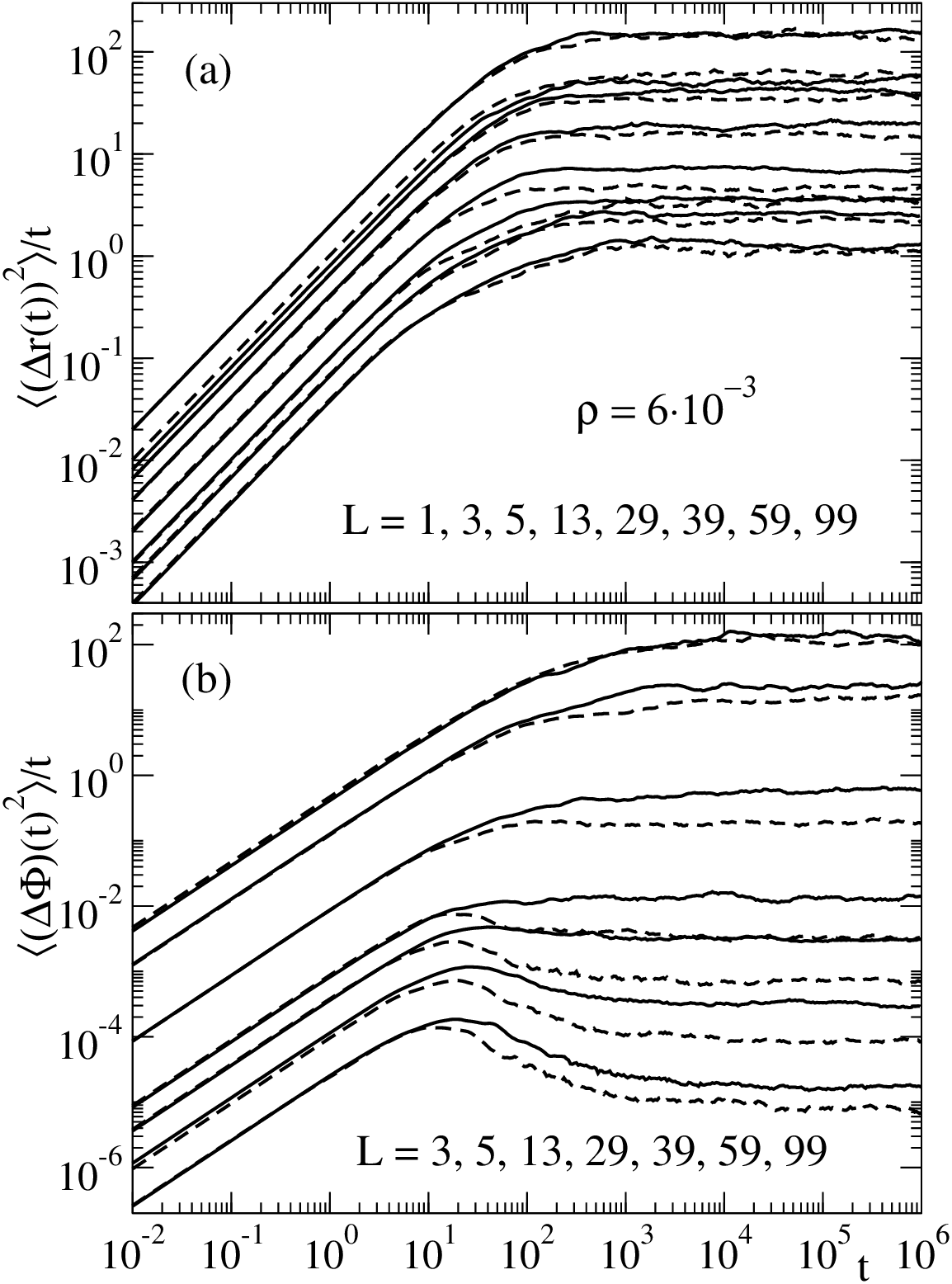}
\caption{Mean-squared displacement of the center-of-mass (a) and mean-squared
angular displacement (b), both divided by the time $t$, for $\rho =6\cdot 10^{-3}$ and
different rod lengths, for the random (solid lines) and glassy (dashed lines)
configuration of the obstacles.}
\end{figure}

No significant differences are observed between the values of
$\langle(\Delta r(t))^{2}\rangle$ for a same rod length in the two
different, random and glassy, configurations. This is more clearly seen
by calculating for each $L$ the ratio $D_{\rm CM}^{\rm ran}/D_{\rm
CM}^{\rm gla}$ between the center-of-mass translational diffusion
constant $D_{\rm CM}$ in both configurations. The latter is calculated
as the long-time limit of $\langle(\Delta r(t))^{2}\rangle/4t$. As
shown in Fig. 2, differences in $D_{\rm CM}$ between the random and the
glassy host medium are less than a factor 1.5. The values of $D_{\rm
CM}$ are systematically lower for the glassy host medium, evidencing a
stronger backscattering for the diffusion, as intuitively expected from
the homogeneous character of this latter configuration  in contrast to
the random one (see Fig. 3), where the presence of holes facilitates
diffusion. It is noteworthy that the maximum difference between the
values of $D_{\rm CM}$ in both configurations takes place for $L \sim
10-20$, i.e., when the rod becomes longer than $d_{\rm nn}$. Thus, while
in the homogeneous glassy host medium the rod will not be able to pass
transversally between two neighboring obstacles, in the inhomogeneous
random configuration this diffusion channel will still be present due to
the presence of holes and will only be suppressed for very long rods.

\begin{figure}
\includegraphics[width=.80\linewidth]{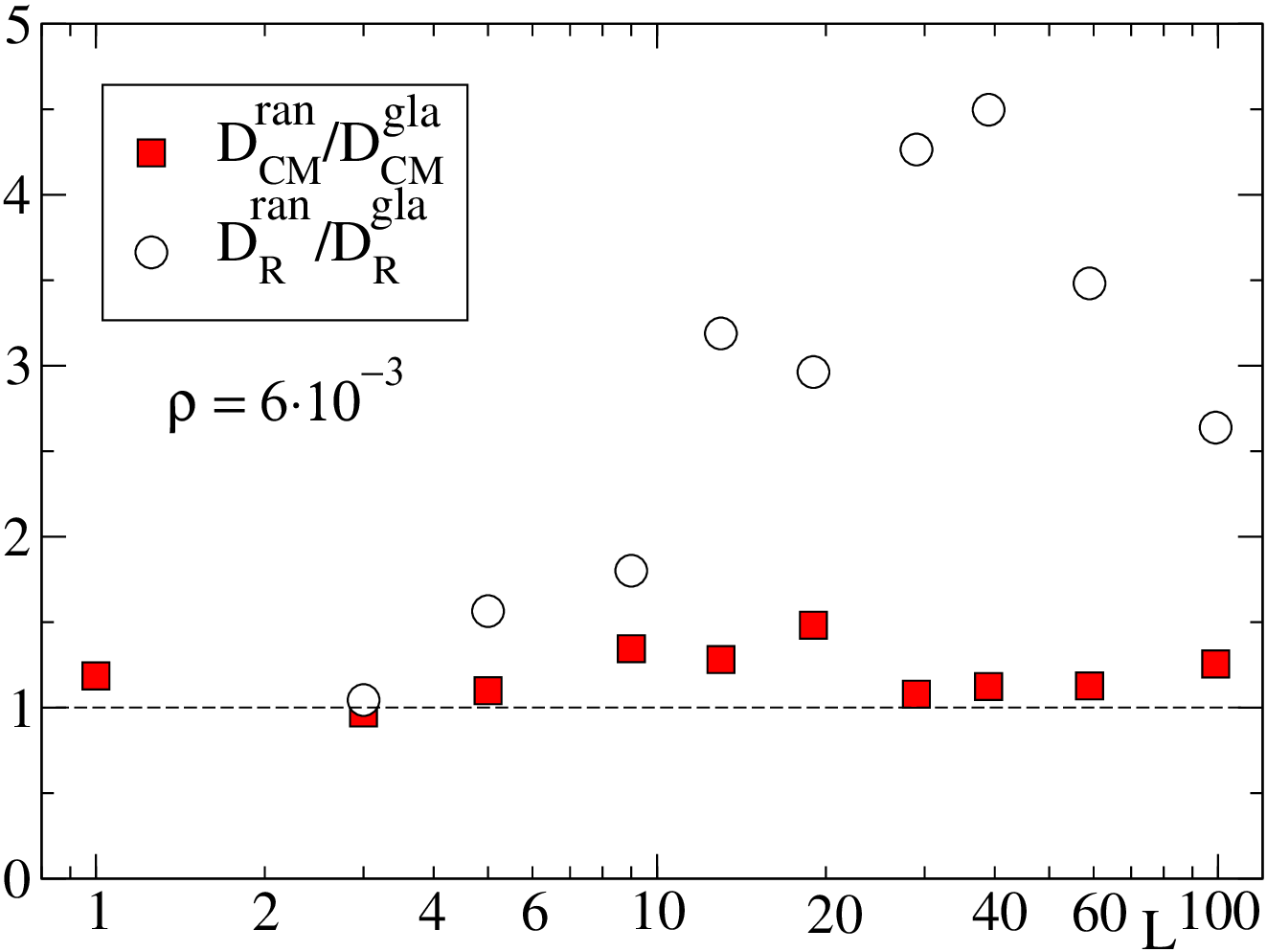}
\newline
\caption{Ratio between the rotational and center-of-mass translational diffusion
constants for the random and the glassy host medium at $\rho = 6\cdot 10^{-3}$
and different rod lengths.\vspace{1.5 cm}}
\end{figure}

\begin{figure}
\includegraphics[width=.80\linewidth]{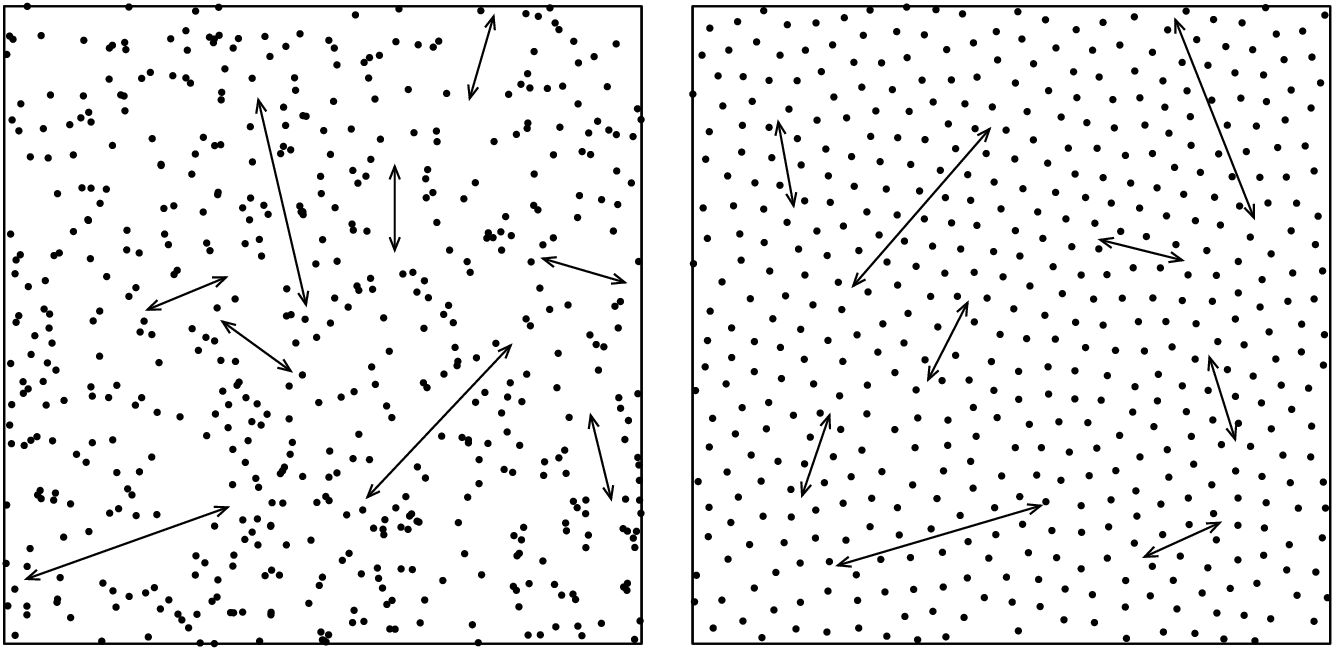}
\newline
\newline
\newline
\caption{Two configurations of the obstacles for a same density $\rho =6\cdot 10^{-3}$.
Left: A random configuration. Right: A homogeneous glassy configuration.
Short and long arrows correspond, respectively, to distances of 40 and 100.}
\end{figure}

Strong differences are observed between both models of the host
medium in the case of the rotational dynamics, as shown in Fig. 1b for
$\langle(\Delta \Phi(t))^{2}\rangle$ and in Fig. 2 for the ratio of the
rotational diffusion constants, $D_{\rm R}^{\rm ran}/D_{\rm R}^{\rm gla}$.
($D_{\rm R}$ is calculated as the long-time limit of $\langle(\Delta
\Phi(t))^{2}\rangle/2t$.) For rod lengths smaller than $d_{\rm nn}$
the rotational dynamics is only weakly sensitive to the configuration
of the medium and the ratio $D_{\rm R}^{\rm ran}/D_{\rm R}^{\rm gla}$
remains close to unity.  However, for rods longer than $d_{\rm nn}$
the latter ratio strongly increases up to a maximum of $\sim 4.5$
for $L \sim 40$.  The position of the maximum can be again rationalized from the
inhomogeneous structure of the random host medium. Short arrows in the
configurations of Fig. 3 for $\rho = 6 \cdot 10 ^{-3}$ indicate distances
of 40. From the comparison between both configurations it is clear that
while in the glassy medium rods of $L \sim 40$ can perform only small
rotations within the tubes formed by the neighboring obstacles, in
the random medium they can go into the holes and thus rotate freely over a larger angle.
Therefore, the long-time angular displacement will grow up more quickly
than for the motion between narrow tubes in the glassy configuration.
It must also be mentioned that even for the largest investigated rod
length, $L = 99$, where motion between tubes also dominates the diffusion
in the random medium, the ratio $D_{\rm R}^{\rm ran}/D_{\rm R}^{\rm gla}$
is still significantly different from unity. Thus, while in the glassy medium the walls
of the tube are formed by uniformly distributed obstacles, in the random
configuration long distances are allowed between some of the neighboring
obstacles forming the tube, leading to additional escaping channels (see
long arrows in Fig. 3).  In principle, only for extremely long rods the ratio
$D_{\rm R}^{\rm ran}/D_{\rm R}^{\rm gla}$ is expected to approach unity.

\section{IV. Correlation functions}

The Mode Coupling Theory (MCT) \cite{19,20,21,22}
makes precise quantitative predictions on the
dynamics of supercooled liquids or dense colloidal systems. In its idealized version,
it predicts a dynamic transition from an ergodic to a non-ergodic phase
at some critical value of the control parameters. These are usually the
temperature or the density, though in principle the MCT formalism can be
generalized to other control parameters. Moreover, it has been recently
tested in a kind of systems very different from liquids or colloids,
such as plastic crystals \cite{23}, suggesting a more universal
character for this theory.  Motivated by this possibilities, we test
some predictions of MCT for the rotational dynamics of the rod with $L$
and $\rho$ as control parameters.
\begin{figure}
\includegraphics[width=.85\linewidth]{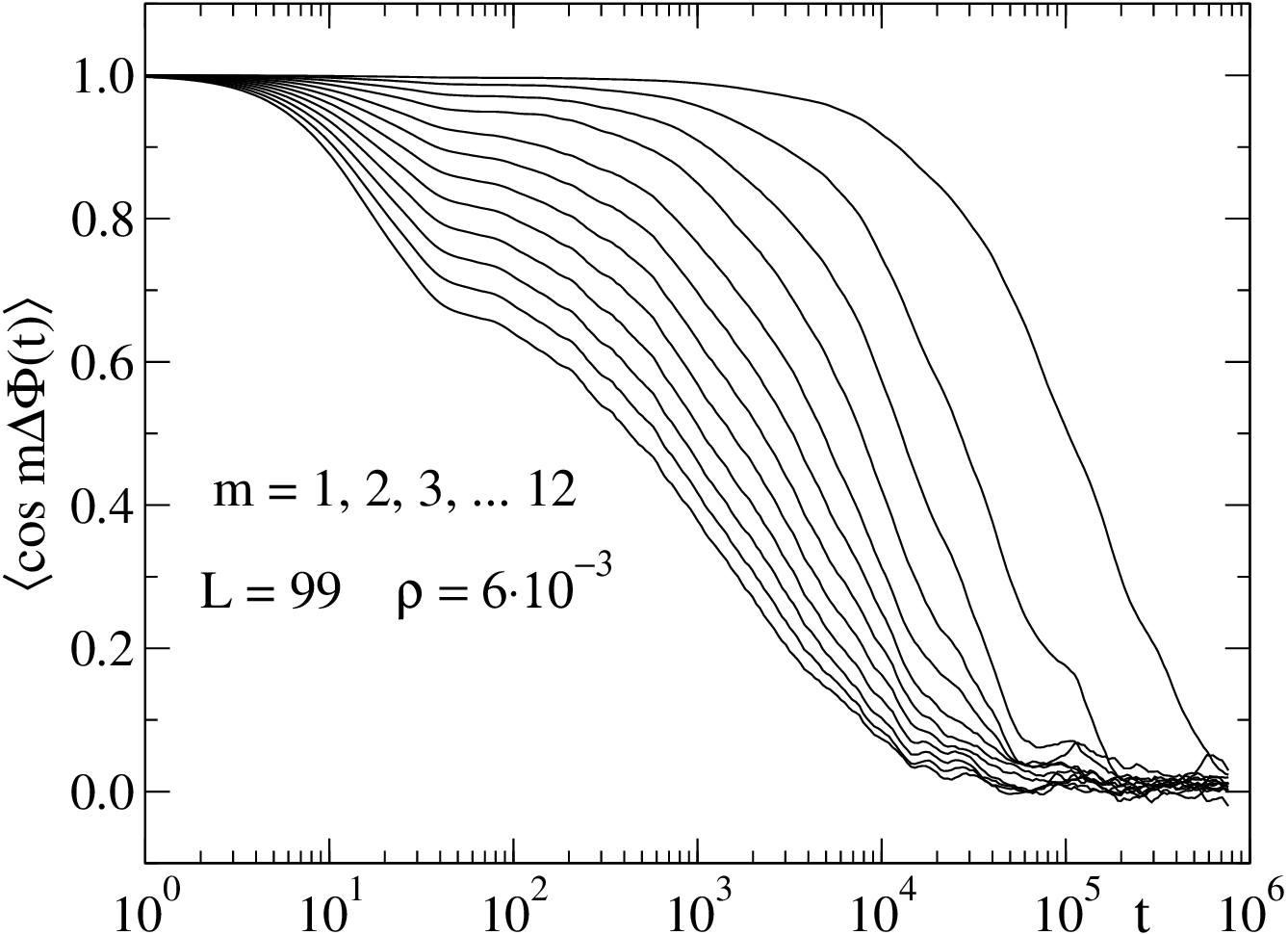}
\newline
\caption{Correlation function $\langle\cos m\Delta\Phi(t)\rangle$ for
$L=99$ and $\rho = 6\cdot 10^{-3}$, with $m =1, 2, 3...12$ (from top to bottom).
Data correspond to the random host medium.}
\end{figure}

According to MCT, upon approaching the critical point, the correlation
function shows a two-step decay. The initial decay corresponds
to the dynamic transition from the ballistic to the caging regime. As
a consequence of the temporary trapping in the cage formed by the
neighboring obstacles, the particle does not lose the memory of its
initial position and the correlation functions decay very slowly, giving rise
to a plateau at intermediate times between the ballistic and the diffusive
regime. The dynamics close to this plateau is usually referred as the $\beta$-relaxation. The
closer the control parameters are to the critical values, the slower
is the mobility of the particles and the longer is the lifetime of the
cage, leading to a longer plateau in the correlation function.  Finally,
at much longer times, the particle escapes from the cage and loses memory
of its initial environment, leading to a second long-time decay of the
correlation function to zero, known as the $\alpha$-relaxation.

Fig. 4 shows the behavior of the angular correlators $\langle\cos
m\Delta\Phi(t)\rangle$ for $m = 1,2,3,...12$ for a density of obstacles
$\rho =6 \cdot 10^{-3}$ and a rod length $L =99$ much longer than $d_{\rm
nn}$. Data are shown for the random host medium, though for the glassy one
they show the same qualitative behavior.  Though it is difficult to
see it for the smallest values of $m$, the plateau is clearly visible
for $m \geq 4$. Thus, long rods will only perform small rotations before
hitting the walls of the tube.  As a consequence, the correlators for small $m$
will decrease only very weakly during the ballistic regime, and hence it
will be difficult to see the subsequent plateau. In contrast to this, for
higher-order correlators, the angle rotated during the ballistic regime
will be amplified by a factor $m$, leading to a stronger decay before the
caging regime and facilitating the observation of the plateau.	The latter
begins to develop around $t \sim 30$.  As can be seen in Fig. 1b for the
curve for $L=99$, this time corresponds to the beginning of the caging
regime for the mean-squared angular displacement, in agreement with the
MCT prediction.
\begin{figure}
  \includegraphics[width=.85\linewidth]{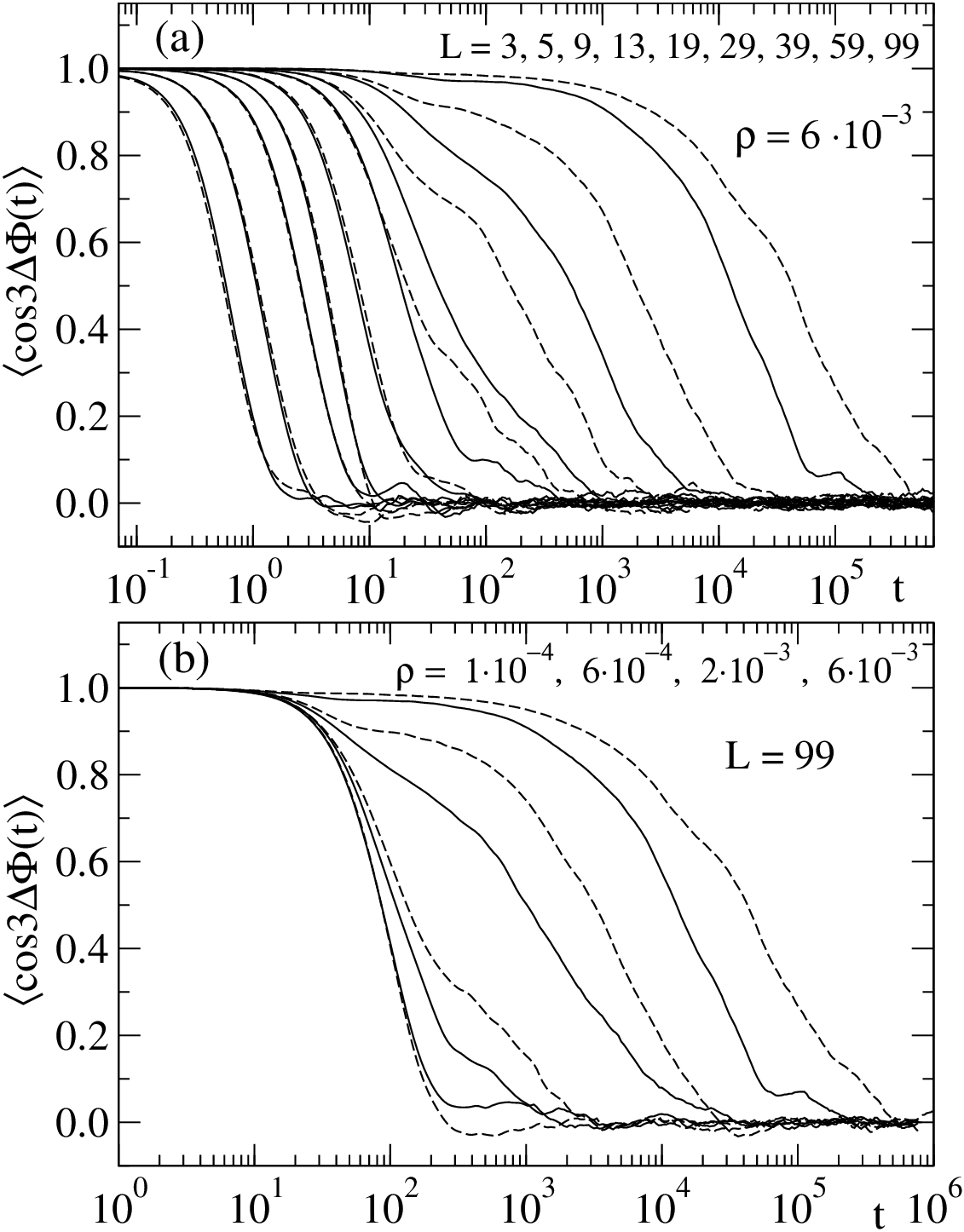}
  \newline
  \caption{(a) Correlation function $\langle\cos3\Delta\Phi(t)\rangle$ for
  $\rho = 6\cdot 10^{-3}$ and different rod lengths; (b) The same function for $L = 99$
  and different densities. Solid and dashed lines correspond, respectively, to the
  random and the glassy configurations of the host medium.
\vspace{0.25 cm}}
\end{figure}

\begin{figure}
  \includegraphics[width=.85\linewidth]{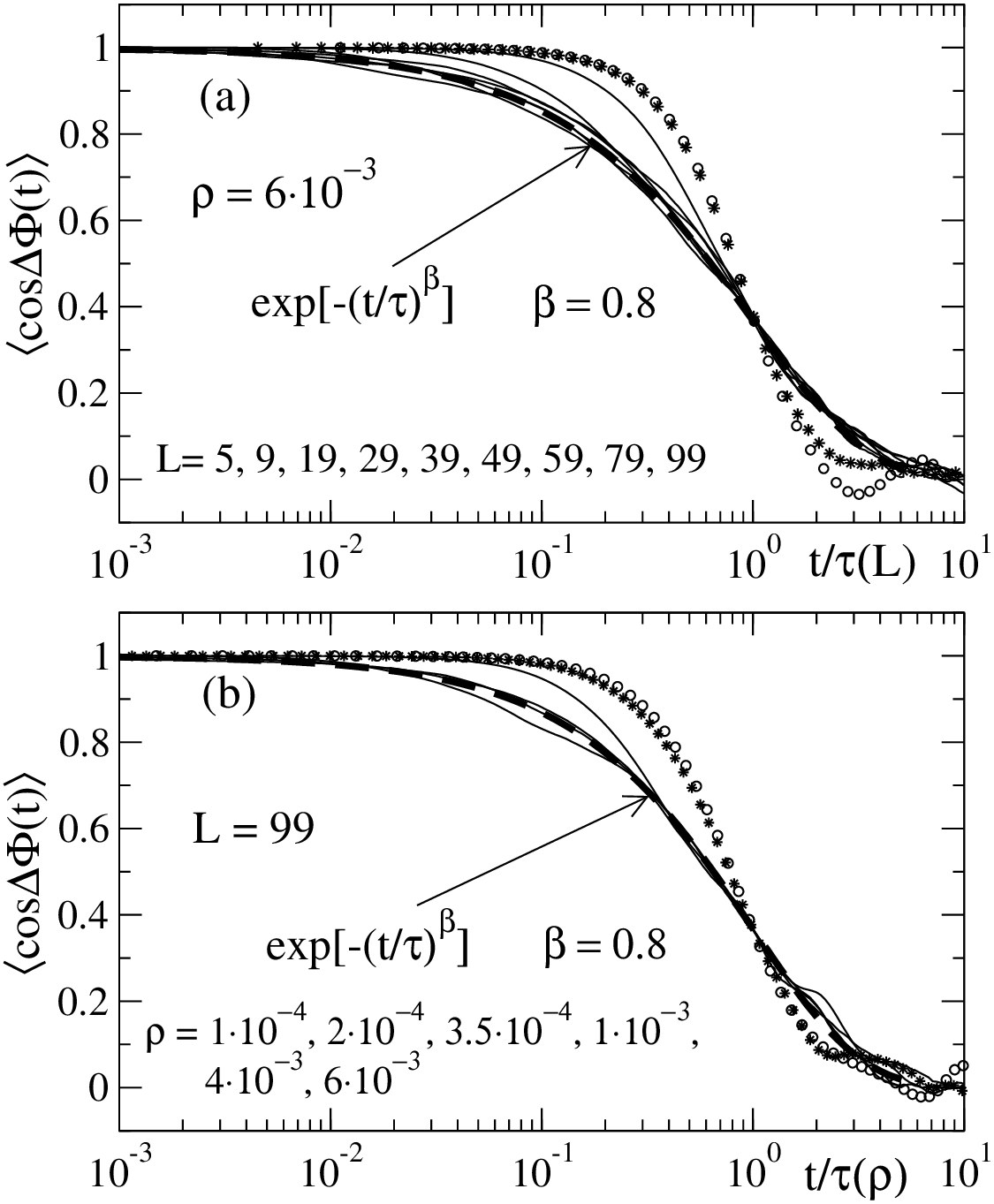}
  \newline
  \caption{(a) Time-length superposition for $\langle\cos\Delta\Phi(t)\rangle$
  at density $\rho = 6\cdot 10^{-3}$; (b) Time-density scaling for the same function
  for length $L = 99$. For the sake of clarity the lowest values of $\rho$ and $L$
  have been represented with points. Dashed lines are fits to KWW functions. The stretching
  exponents are indicated in the figure. Data correspond to the random host medium.}
\end{figure}

Fig. 5 shows for $m=3$, and for the two models of host medium, how
the plateau develops with increasing rod length and density of obstacles.
While for values of $L$ below $d_{\rm nn}$, the correlators show a simple
decay and within the error bar show no difference between both models of
the host medium, a clear difference is observed for longer rods. Thus,
for the glassy medium rotational relaxation times become about a decade
larger and the plateau is significantly higher than in the random
configuration, in agreement with the results presented in Section III
for the mean-squared angular displacement.

Another important prediction of the MCT is the so called ``second
universality'': given a correlator $g(t,\zeta)$, with $\zeta$ a
control parameter,  then one has that, in the time scale of the
$\alpha$-regime, the correlator follows a scaling law $g(t,\zeta) =
\tilde{g}(t/\tau(\zeta))$, where $\tau(\zeta)$ is the $\zeta$-dependence
of the relaxation time $\tau$ of the $\alpha$-regime for such a
correlator, and $\tilde{g}$ is a master function.  
The relaxation time is in practice defined as the time
where the correlator decays to an arbitrary but small fraction of its
initial value, or is obtained from fitting the $\alpha$-decay of the
correlator to a Kohlrausch-Williams-Watts (KWW) function,
$\exp(-(t/\tau)^{\beta})$, widely used in the analysis of relaxations
in complex systems \cite{24}.  In most of the experimental situations
the relevant control parameter is the temperature and for that reason,
the second universality is often referred as the ``time-temperature
superposition principle''.  Now we test the existence of a time-length
and a time-density superposition principle for the angular correlators of
the rod.  Fig.~6 shows the correlators $\langle\cos\Delta\Phi(t)\rangle$
as a function of the scaled times $t/\tau(L)$ for constant density $\rho
=6 \cdot 10^{-3}$, and $t/\tau(\rho)$ for constant rod length $L=99$.
Data are shown for the random host medium. The relaxation times $\tau$
have been defined as $\langle\cos\Delta\Phi(\tau)\rangle = 1/e$.  Apart
from the trivial scaling in the limit of short rods and low densities,
a good superposition to a master curve is obtained for larger values of
$L$ and $\rho$, confirming the second universality of the MCT for this
system. The master curve can be well reproduced by a KWW function with a
stretching exponent $\beta = 0.8$, as shown in Fig. 6.  Analogous results,
with a  very similar stretching exponent, are obtained for the glassy host medium.

\section{V. Non-Gaussian effects}

In Einstein's random walk model for diffusion, particles move under
the effect of collisions with the others.  When a particle undergoes a
collision, it changes its direction randomly, and completely loses the
memory of its previous history.  When the time and spatial observational
scales are much larger than the characteristic time and the
mean free path between collisions, the van Hove self-correlation function,
i.e., the time and spatial probability distribution $G_{\rm s}(r,t)$
of a particle initially located at the origin, is given by a Gaussian
function \cite{25}.  This functional form is exact for an ideal gas and
for an harmonic crystal.  It is also valid in the limit of short times,
where atoms behave as free particles. When the system is observed at time
and length scales comparable to those characteristic of the collisions,
the possible intrinsic dynamic heterogeneities of the system will be
reflected by strong deviations from Gaussianity in $G_{\rm s}(r,t)$.
Such deviations are usually quantified by the so-called second-order
non-Gaussian parameter $\alpha_{2}(t)$, which in two-dimensions is defined
as $\alpha_{2}(t) = [\langle (\Delta r(t))^{4}\rangle/2\langle (\Delta
r(t))^{2}\rangle^{2}]-1$.  For a Gaussian function in two dimensions,
$\alpha_{2}(t)=0$, while deviations from Gaussianity result in finite
values of $\alpha_{2}(t)$.

Our previous investigation \cite{10} on the non-Gaussian parameter
in the random host medium revealed some features similar to those
observed in supercooled liquids or dense colloids, such as the development
of a peak (see also Fig.~7), which grows with increasing rod length
-in analogy to decreasing temperature in supercooled liquids or
increasing density in colloids.  As also observed in these latter systems
\cite{11,12,13,14,15,16,17,18}, the region around the maximum of the peak
corresponds to the time interval corresponding to the end of the caging and
the beginning of the long-time diffusive regime.  Thus, the breaking
of the cage leads to a finite probability of jumps much longer that
the size of the cage, resulting in a strongly heterogeneous dynamics at
that intermediate time and spatial scale, which is reflected by a peak
in the non-Gaussian parameter.
\begin{figure}
  \includegraphics[width=.87\linewidth]{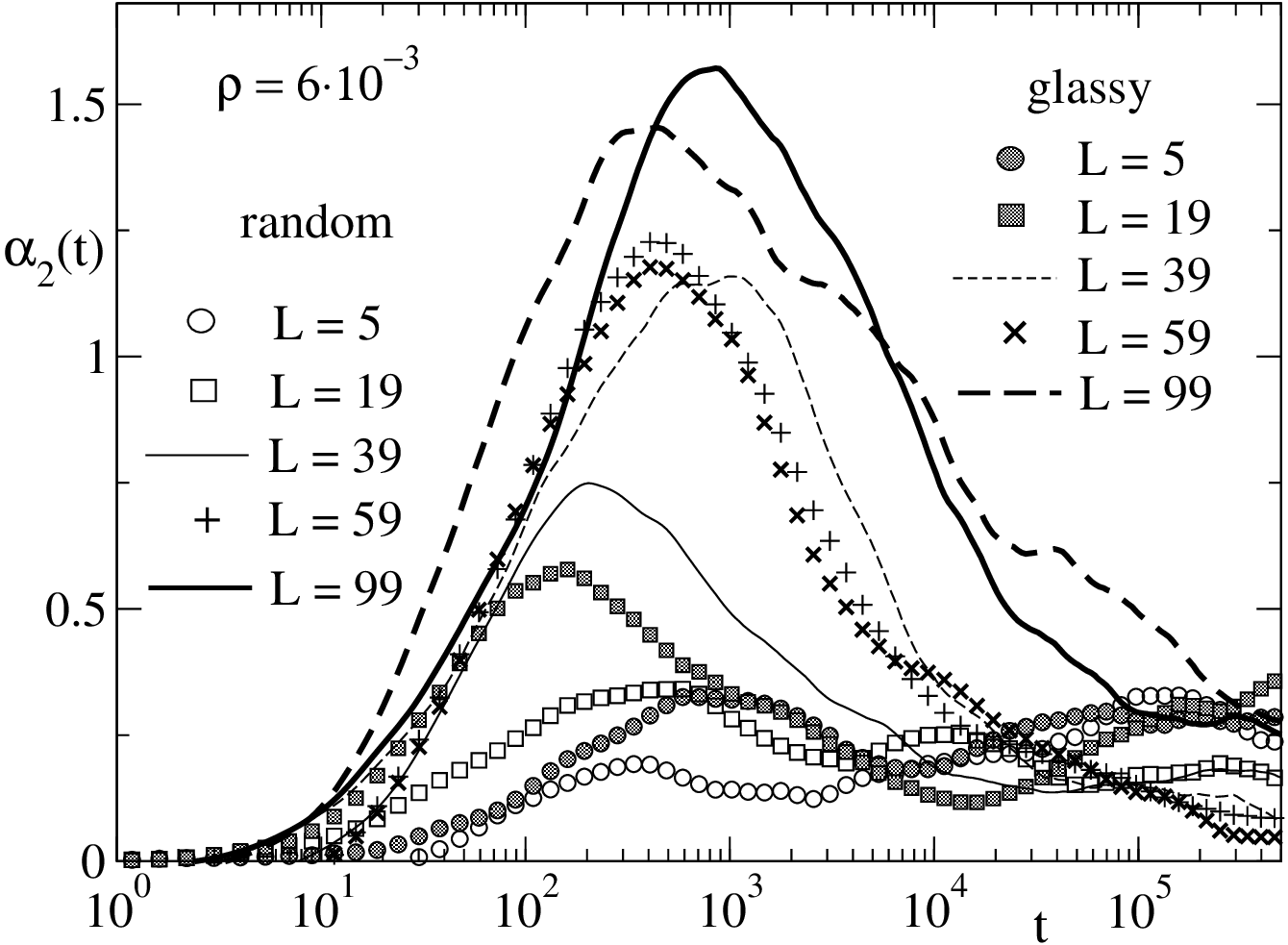}
  \newline
  \caption{Time-dependence of the non-Gaussian parameter $\alpha_2(t)$ for different values
of $L$ and for the random and the glassy host medium
  at density $\rho = 6\cdot 10^{-3}$.}
\end{figure}

Another interesting result was the observation that, in particular for long rods, 
in the time span
of the simulation, i.e. for time scales that are much longer than the typical relaxation
time of the system, $\alpha_{2}(t)$ did not decay to zero but remained
finite~\cite{10}. Thus, the long-time dynamics
is still significantly non-Gaussian, i.e. heterogeneous, at the spatial
scale -much larger than the rod length- covered by the simulation. In order to
investigate the possibility of a relation of this large-scale non-Gaussian
dynamics with the presence of holes in the random structure of obstacles,
that might lead to a finite probability of jumps much longer than the
average, as pointed out in Ref.[10], we have calculated $\alpha_{2}(t)$
for the dynamics of the rod in the glassy host medium. Fig.~7 shows
a comparison between the non-Gaussian parameter in both structures
for several rod lengths and density $\rho =6 \cdot 10^{-3}$. While in
the limit of short and long rods $\alpha_{2}(t)$ takes similar values
for both configurations in all the time window investigated, a much
more pronounced peak is observed at intermediate rod lengths for the
glassy configuration.  This result can be rationalized by the absence
of holes in the homogeneous glassy host medium.  Thus in the latter,
jumps that lead to an escape from the cage formed by the neighboring obstacles will be
necessarily long, since long rods will be confined in tubes which they will
be able to leave only by making a long longitudinal motion. 
In contrast to this, the above mentioned inhomogeneous nature of the
tube walls in the random structure will allow the rod to escape the cage
also by shorter jumps, resulting in a less heterogenous caging regime. Only
rods much longer than the hole size will need long jumps to escape from
the tubes in the random host medium and the caging regime will become as
heterogeneous as in the glassy host medium, as evidenced by the similar
peak heights in Fig.~7 for $L = 59$ and 99.

Concerning the long-time dynamics, the non-Gaussian parameter also
remains finite for the glassy host medium, and no important differences
with the random configuration are observed.  Therefore, contrary to
what was previously pointed out~\cite{10}, the non-Gaussianity observed at large
length scales for the long-time dynamics in the inhomogeneous random
host medium is also present in the homogeneous glassy one, and is not
related to the presence of holes in the configuration of obstacles,
the latter having effects only in the intermediate caging regime.

\begin{figure}
  \includegraphics[width=.85\linewidth]{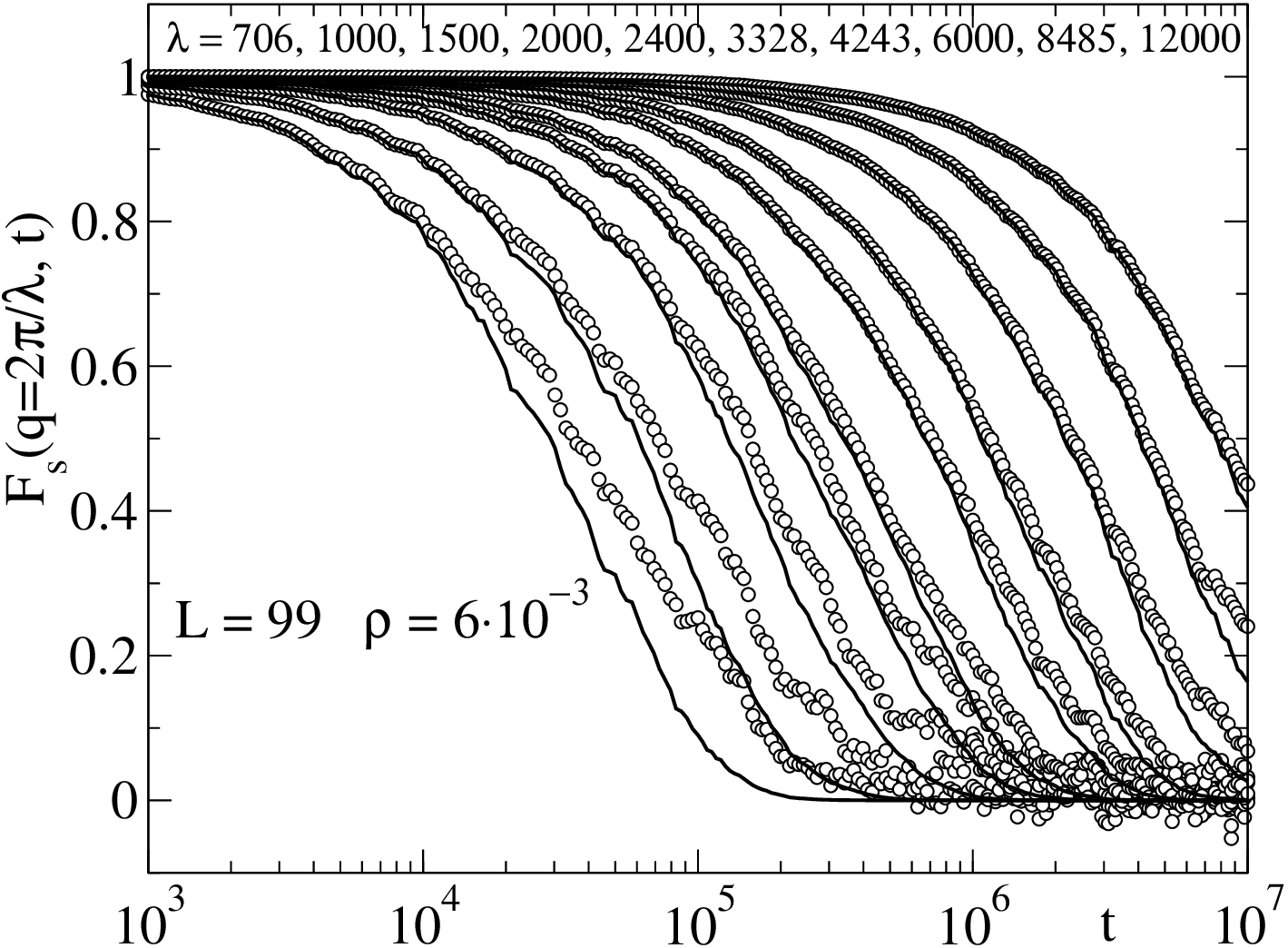}
  \newline
  \caption{Intermediate incoherent scattering function $F_{\rm s}(q=2\pi/\lambda,t)$ at different
  wavelengths $\lambda$ for $\rho = 6\cdot 10^{-3}$ and $L=99$. Points correspond to simulation data in
  the random host medium. Lines are the curves in Gaussian approximation as calculated from
  the center-of-mass mean-squared displacement obtained from the simulation (see text).}
\end{figure}

Another way of visualize the non-Gaussian effects at large length
scales is obtained by representing the intermediate incoherent
scattering  function $F_{\rm s}(q,t)=\langle\exp(-i{\bf q}\cdot\Delta{\bf
r}(t))\rangle$, at very long wavelengths $\lambda = 2\pi/q$. Brackets
denote ensemble and angular average.  Fig. 8 shows this latter function
in the random host medium for $L =99$ and $\rho = 6\cdot 10^{-3}$,
at different (long) wavelengths. (Note that the simulations have been extended one
order of magnitude respect to those presented in Ref.~\cite{10}.) The presence of
non-Gaussian effects are made clear by comparing the different curves
with those corresponding to the Gaussian case in two-dimensions
\cite{25}, $\exp[-\langle(\Delta r(t))^{2}\rangle q^{2}/4]$, where we
use the center-of-mass mean-squared displacement $\langle (\Delta
r(t))^{2}\rangle$ calculated from the simulations.  From this comparison
it is clear that significant non-Gaussian effects are still present at
length scales of at least $\sim 9000$, i.e. two orders of magnitude larger than the
rod length.  Whether this result indicates that the diffusion of a rod
in a disordered array of obstacles is actually a Poisson process at any
length scale is an open question.

\section{VI. Conclusions}

By means of molecular dynamics simulations, we have compared the dynamics
of a rigid rod in two models of a 2d-disordered static host medium:
a random configuration of soft disks and another ``glassy'' one obtained
from freezing a liquid of soft disks in equilibrium. While the former
is characterized by the presence of big holes and clusters of close
obstacles, the latter presents an homogeneous structure.  No significant
differences have been observed in the translational dynamics of the rod
center-of-mass. However, rotations are much more hindered in
the glassy host medium.

Angular correlation functions have been calculated for a wide range of
rod length and density of obstacles. In agreement with the predictions
of the Mode Coupling Theory, these functions show a plateau at the time
scale of the caging regime, and follow a time-length and a time-density
scaling for the long-time dynamics.

Strong non-Gaussian behavior has been observed at large length scales,
though no significant differences are evidenced between the random and
the glassy configuration of the obstacles. This result suggests that
the long-time translational dynamics is not controlled by the presence
of inhomogeneities in the host medium.

We thank E. Frey for useful discussions.  A.J.M acknowledges a
postdoctoral grant from the Basque Government. Part of this work was
supported by the European Community's Human Potential Program under
contract HPRN-CT-2002-00307, DYGLAGEMEM.

\end{document}